\documentclass[twocolumn,showpacs,aps,amsmath,amssymb,prl,superscriptaddress]{revtex4-1}
\usepackage{amsfonts}
\usepackage{amsmath}
\usepackage{bm}
\usepackage{graphicx}

\newcommand{\be}{\begin{equation*}}
\newcommand{\ee}{\end{equation*}}
\newcommand{\bea}{\begin{eqnarray}}
\newcommand{\eea}{\end{eqnarray}}
\newcommand{\Fig}[1]{Fig.~\ref{#1}}

\newcommand{\Eq}[1]{Eq.~(\ref{#1})}

\newcommand{\w}{\omega}
\newcommand{\ep}{\epsilon}

\begin{document}

\title{Thermopower of few-electron quantum dots with Kondo correlations}

\author{LvZhou Ye}  
\affiliation{Department of Chemical Physics, University of Science and Technology of China, Hefei, Anhui 230026, China}

\author{Dong Hou}
\affiliation{Hefei National Laboratory for Physical Sciences at the
Microscale, University of Science and Technology of China, Hefei, Anhui
230026, China}
\affiliation{Department of Chemistry, Hong Kong University of Science
and Technology, Hong Kong, China}

\author{Rulin Wang}
\affiliation{Department of Physics, University of Science and Technology of China, Hefei, Anhui 230026, China}

\author{Xiao Zheng} \email{xz58@ustc.edu.cn}
\affiliation{Department of Chemical Physics, University of Science and Technology of China, Hefei, Anhui 230026, China}
\affiliation{Hefei National Laboratory for Physical Sciences at the
Microscale, University of Science and Technology of China, Hefei, Anhui
230026, China}

\author{YiJing Yan} \email{yyan@ust.hk}
\affiliation{Department of Chemical Physics, University of Science and Technology of China, Hefei, Anhui 230026, China}
\affiliation{Hefei National Laboratory for Physical Sciences at the
Microscale, University of Science and Technology of China, Hefei, Anhui
230026, China}
\affiliation{Department of Chemistry, Hong Kong University of Science
and Technology, Hong Kong, China}

\date{\today}

\begin{abstract}

  The thermopower of few-electron quantum dots with Kondo correlations is investigated
via a hierarchial equations of motion approach. The thermopower is determined
by the line shape of spectral function within a narrow energy window defined by temperature.
Based on calculations and analyses on single-level and two-level Anderson impurity models,
the underlying relations between thermopower and various types of electron correlations are
elaborated. In particular, an unconventional sign reversal behavior is predicted for
quantum dots with a suitable inter-level spacing. The new feature highlights the significance of
multi-level effects and their interplay with Kondo correlations. Our finding and
understanding may lead to novel thermoelectric applications of quantum dots.

\end{abstract}


\maketitle

Thermopower is one of the fundamental thermoelectric properties.
It measures the thermovoltage ($V_T$) induced by a temperature gradient ($\Delta T$).
Materials with a large thermopower are potentially very useful for a variety of applications,
such as electronic refrigeration \cite{Dub11131}, thermoelectric conversion \cite{Yad113555},
and on-chip cooling \cite{Kan10648}.
Recently, the thermopower of nanostructured materials,
such as quantum wires, quantum dots (QDs), and molecular junctions, have been found significantly larger than the prediction of the Wiedemann-Franz law \cite{Sha11399}.
Experimental measurements \cite{Sma03256801,God992927,Yu051842,Sch05176602,Sch07041301,Sch08083016,%
Hoc08163,Gos09026602,Bal11569,Jaw12210,Red071568,Sha11399}
and theoretical calculations \cite{Lun06256802,Dub09081302,Cos10235127,Liu10245323,%
Rej12085117,Mun13016601,Aze12075303,Wan10057202}
on the thermopower have been extensive in the literature.
%
%
In particular, it has been found that in QDs the phonon contribution to thermoelectric properties is greatly suppressed \cite{Man114679}. Therefore, thermopower can be deemed as an intrinsic electronic property of QDs and is very sensitive to the details of electronic structure.

The thermopower of a QD with few electrons is tunable by varying the discrete energy levels with a gate voltage.
This has been realized by Scheibner \emph{et al.} on a QD of 20 to 40 electrons \cite{Sch05176602}.
They have observed that the line shapes of thermopower in the Kondo regime are qualitatively
different from those in the Coulomb blockade regime \cite{Sch05176602}.
However, the predominant effects leading to the observation have remained largely unexplored. Although ``it will be interesting to look for these effects in QDs in the very-few-electron limit'' \cite{Sch07041301,Cio0016315}, relevant studies have remained rather scarce.

The major challenge for theoretical studies is the accurate characterization of Kondo correlations.
A number of approaches have been employed to investigate properties of strongly correlated QDs.
These include the numerical renormalization group (NRG) method \cite{Wil75773,Bul08395}, the Bethe Ansatz \cite{And83331},
the quantum Monte Carlo method \cite{Hir862521,Gul11349}, the exact diagonalization \cite{Caf941545,Dag94763}, and
the hierarchial equations of motion (HEOM) approach \cite{Jin08234703,Zhe09164708,Zhe121129,Li12266403}.
%
%
Using the NRG method, Costi \emph{et al.}~\cite{Cos10235127} studied the temperature and gate voltage dependence of
thermopower for a single-level Anderson impurity model.
Since a real QD usually consists of multiple levels, interactions among electrons at different levels are expected to play nontrivial roles. Therefore,
it is highly desirable to have the multi-level effects included in a theoretical model.

In this letter, we adopt the HEOM approach, an accurate and universal formalism for quantum open systems  \cite{Jin08234703} to study both single-level and two-level QDs. The HEOM approach has been used to characterize various equilibrium and nonequilibrium properties of strongly correlated quantum impurity systems \cite{Zhe08184112,Li12266403,Wang2013},
including the dynamic Coulomb blockade \cite{Zhe08093016} and dynamic Kondo transitions \cite{Zhe09164708,Zheng2013}.

Thermopower is usually measured as the Seebeck coefficient $S\equiv V_T/\Delta T$ in the vanishing current limit ($I=0$).
%
%
A rigorous way to calculate $S$ is to search for the bias voltage $\Delta V$ which cancels exactly the $V_T$ induced by the given $\Delta T$ \cite{Wie10165334}.
In the linear regime where both $\Delta V$ and $\Delta T$ are sufficiently small, an equivalent and often more convenient way is $S=L_T/G$,
since $I=G\Delta V+L_T\Delta T$ \cite{Tur02115332}. Here, $G$ is the conductance at equilibrium and $L_T$ is a coefficient measuring the electric current driven by temperature gradient.
Conventionally $G$ and $L_T$ are obtained by calculating some related equilibrium quantities \cite{Mei913048}.
The HEOM approach admits the above both ways, and the resulting $S$ are affirmed to be numerically equivalent in the linear regime \cite{Sup1}.

In the framework of HEOM, the hierarchy needs to be truncated at a certain level $L$ to close the equations. The results are quantitatively accurate as long as they converge with respect to $L$. Usually a higher $L$ (computationally more costly) is required to achieve the convergence at a lower temperature $T$. In practice, a low $L$ is often found sufficient at a finite $T$. The results presented in this letter are verified as converged at $L = 4$ unless otherwise specified.

\begin{figure}
  \includegraphics[width=\columnwidth]{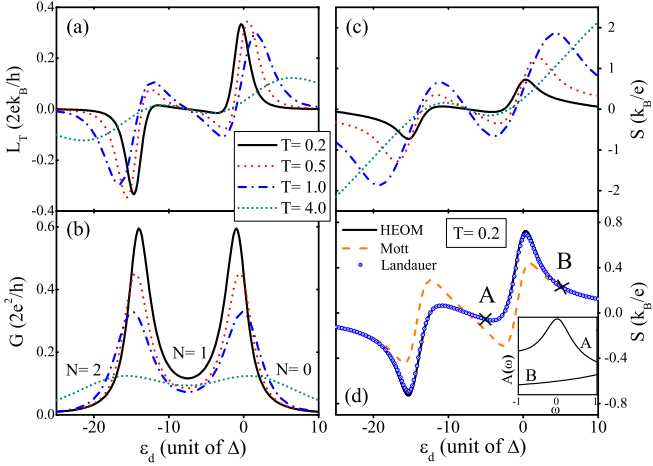}
  \caption{(Color online). Variation of (a) $L_T$, (b) $G$, and (c) $S$ versus $\epsilon_d$ for a  single-level QD at various temperatures. Other parameters  (in unit of $\Delta$) are $U=15$ and $W=30$.
  (d) A comparison of $S_{\mathrm{HEOM}}$, $S_{\mathrm{Mott}}$ and $S_{\mathrm{Landauer}}$ at $T = 0.2\,\Delta$.
  The inset in (d) shows $A(\w)$ around $\mu$ for
  $\epsilon_d=-5\Delta$ (marked by A) and $\epsilon_d=5\Delta$ (marked by B), respectively.
  }\label{fig1}
\end{figure}


We first examine the thermopower of a single-level QD represented by the Hamiltonian of $H = H_{\rm dot} + H_{\rm lead} + H_{\rm coup}$.
Here, $H_{\rm dot}=\epsilon_d (\hat n_\uparrow + \hat n_\downarrow ) +U \hat n_\uparrow \hat n_\downarrow$ is the dot Hamiltonian, where $\hat n_s = \hat a^\dagger_s \hat a_s$, $\hat a^\dagger_s$ ($\hat a_s$) creates (annihilates) an electron of spin-$s$ on the dot level of energy $\epsilon_d$, and $U$ is the Coulomb repulsion energy.
$H_{\rm lead} = \sum_{\alpha k} \epsilon_{\alpha k}\, \hat d^\dagger_{\alpha k}\hat d_{\alpha k}$ describes the two noninteracting leads, where $\hat d^\dagger_{\alpha k}$ ($\hat d_{\alpha k}$) creates (annihilates) an electron on lead-$\alpha$ state $|k\rangle$ of energy $\epsilon_{\alpha k}$.
$H_{\rm{coup}} = \sum_{\alpha k} t_{\alpha k}\, \hat a^\dagger_{s}\, \hat d_{\alpha k} + \rm{H.c.}$ represents dot-lead couplings, with $t_{\alpha k}$ being the coupling strength between the dot level and the lead-$\alpha$ state $|k\rangle$.
The lead information enters the HEOM only via the hybridization functions,
$\Delta_{\alpha}(\omega) \equiv \pi \sum_{\alpha k} |t_{\alpha k}|^2 \delta(\omega-\epsilon_{\alpha k})$,
which assumes a form of
$\Delta_{\alpha}(\omega)=\frac{1}{2}\Delta /[(\omega-\mu_\alpha)^2/W^2 + 1]$.
Here, $\Delta$ is the effective coupling and $W$ is the band width, which are set as identical for both leads; and $\mu_\alpha$ is the chemical potential of lead-$\alpha$.

Figures\,\ref{fig1}(a)-(c) depict the calculated $L_T$, $G$, and $S$ versus the level energy $\epsilon_d$ at various temperatures, respectively.
As shown in \Fig{fig1}(b), in the valley where the electron occupation $N$ is around $1$, the conductance increases with the decreasing $T$ at $T < \Delta$. This clearly indicates the presence of Kondo resonance
\cite{Gol98156,Cro98540}.
The electron-hole (e-h) symmetry point is $\epsilon^{\rm eh}=-U/2=-7.5\Delta$, \emph{i.e.}, at the center of the Kondo valley. At $\epsilon_d=\epsilon^{\rm eh}$, the thermally induced currents carried by electrons and by holes cancel out exactly, leading to the zero $L_T$ at any $T$.
The line shape of $L_T(\epsilon_d)$ is anti-symmetric with respect to $\epsilon_d=\epsilon^{\rm eh}$, while $G(\epsilon_d)$ is symmetric. Consequently, $S$ is anti-symmetric with respect to the e-h point.
As displayed in \Fig{fig1}(c), for all $T$ studied, $S$ always reverses its sign (from negative to positive) as $\ep_d$ increases from
$\ep^{\rm eh}$ towards the $N=0$ region. This agrees with previous NRG results at similar temperatures \cite{Cos10235127}.

A semi-classical Mott relation is often used for $S$
\cite{Cut691336}:
\begin{equation} \label{mott}
  S_{\mathrm{Mott}}=-\frac{\pi^2}{3}\frac{k^2_{\mathrm{B}}T}{e}\left.\frac{\partial \ln G(\epsilon)}{\partial\epsilon}\right|_{\epsilon=\mu}.
\end{equation}
As shown in \Fig{fig1}(d), $S_{\rm Mott}$ fails
to reproduce the correct line shape of $S$ in the Kondo regime.
In contrast, a Landauer-like formula \cite{Lun053879} recovers the HEOM calculated $S$ quantitatively:
%
\begin{equation} \label{landauer}
  S_{\rm Landauer} = \frac{1}{eT}\frac{\int d\omega \, (\omega-\mu)f'(\w)A(\omega)} {\int d\omega \, f'(\w)A(\omega)}.
\end{equation}
Here, $A(\omega)$ is the equilibrium spectral function of QD, $f(\omega)$ is the Fermi function, and $f'(\w) \equiv \frac{\partial f(\w)}{\partial \w}$. Equation\,\eqref{landauer} relates $S$ to $A(\omega)$ within an energy window centered at the chemical potential $\mu$. The size of window is determined by the width of $|(\omega-\mu)f'(\w)|$ \cite{Sup1}. Apparently, $S$ is positive (negative) if $A(\omega)$ is overall larger in the right (left) half of the window.
It is known that \cite{Dub11131}, only when $A(\w)$ varies slowly within the energy window, can $S_{\rm Mott}$ of \Eq{mott} be recovered from \Eq{landauer} \cite{Sup1}.

The inset of \Fig{fig1}(d) exhibits $A(\w)$ in the vicinity of $\mu \equiv 0$ as $\epsilon_d$ is tuned from within the Kondo regime (marked by A) to a non-Kondo regime (marked by B). At point A, the Kondo resonance gives a prominent peak centered at $\w = \mu$. Since $A(\omega)$ at $\w < \mu$ is overall large than that at $\w > \mu$, $S$ is negative at A. At point B, the Kondo resonance is absent, and $A(\w)$ varies rather slowly with $\w$. Consequently, $S_{\rm Mott}$ is found to agree well with $S_{\rm Landauer}$, and both are of positive values.

\begin{figure}
  \includegraphics[width=\columnwidth]{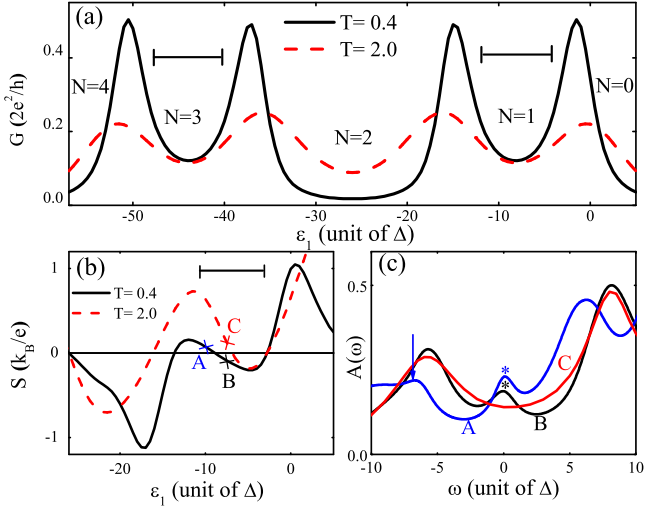}
  \caption{(Color online). (a) $G$, and (b) $S$ versus $\epsilon_1$ for $\delta\epsilon=7\Delta$ at high and low temperatures.
  (c) $A(\w)$ at different $\epsilon_1$ or $T$. Each line in (c) corresponds to a point marked by a cross in (b). The horizontal bars in (a) and (b) indicate the Kondo regimes. The asterisks and vertical arrow in (c) indicate the positions of Kondo peaks and side peak, respectively. Other parameters  (in unit of $\Delta$) are $U = 15$ and $W = 60$.
 }\label{fig2}
\end{figure}

The sign reversal of $S$ has been observed in experiment by Scheibner \emph{et al.} \cite{Sch05176602} at temperatures comparable to \Fig{fig1}. They have also found that, as $T$ is lowered further, $S$ retains positivity for a wide range of $\epsilon_d$ \cite{Sch05176602}. This has been attributed to Kondo correlations \cite{Cos10235127}.
We emphasize that the preservation of the positivity of $S$ is not universal, as Kondo resonance is not the only affecting factor.
Real QDs involve more than one energy levels, especially in the presence of a tunable gate voltage. It is thus intriguing that how the multi-level feature would affect the $S$ of QDs.
Moreover, the e-h point of a single-level QD locates within the Kondo valley, and the e-h symmetry enforces $S = 0$ at its center. This is no longer the case for a multi-level QD, for which the situation is more complex for $S$ in the Kondo regime.

We then consider a two-level QD of
\begin{equation}
H_{\mathrm{dot}}= \sum_{i,s}\epsilon_{i}\, \hat n_{is}
+ U \sum_{i} \hat n_{i\uparrow} \, \hat n_{i\downarrow}
+ U \sum_{s,s'}\hat n_{1s} \, \hat n_{2s'},
\end{equation}
where $i=1,2$ labels the levels. The intra- and inter-level Coulomb interactions assume the same $U$.
For a typical semiconducting QD, the inter-level spacing $\delta \epsilon = \epsilon_{\mathrm{2}}-\epsilon_{\mathrm{1}}$ is an order of magnitude smaller than $U$ \cite{Cro98540}, and it can be tuned experimentally by controlling the size of QD.
Limited by computational resources, an $L=3$ truncation is adopted for calculations on two-level QDs. While the Kondo resonance at the lowest temperature considered ($T=0.4\Delta$) is slightly overestimated, all other quantities are converged \cite{Sup1}, and all conclusions remain valid.

Figure\,\ref{fig2}(a) depicts $G$ versus $\epsilon_1$ for a large level spacing of $\delta\epsilon=7\Delta$ at low ($T=0.4\Delta$) and high ($T=2\Delta$) temperatures.
Each of the valleys corresponds to an integer $N$. At both $N=1$ and $N=3$ valleys, the low-$T$ conductance is larger than the high-$T$ counterpart, suggesting the presence of Kondo resonance at the low $T$.
The corresponding $S$ versus $\epsilon_1$ are shown in \Fig{fig2}(b).
The e-h symmetry point is now at $\epsilon^{\rm eh}=-(3U+\delta\epsilon)/2$.
For clarity, only the $\epsilon_1 > \epsilon^{\rm eh}$ half of $S(\epsilon_1)$ is displayed, and the other half can be obtained from the anti-symmetry.
Clearly, the line shapes of $S$ always exhibit a sign reversal within the Kondo regime.

To understand the sign reversal behavior, the $A(\w)$ for various values of $\epsilon_1$ are plotted in \Fig{fig2}(c).
At the high temperature of $T = 2\Delta$ and $\epsilon_1 = -7.5\Delta$ (line C), the energy window relevant to \Eq{landauer} is so wide that it involves the nearby Hubbard peaks. The higher peak at $\w > \mu$ leads to a positive $S$ at the point C in \Fig{fig2}(b).
As $T$ decreases to $0.4\Delta$ while $\epsilon_1$ is unchanged (line B), a Kondo peak emerges at $\w = \mu$, and meanwhile, the energy window becomes much narrower. The left side of the Kondo peak is higher, leading to a negative $S$ at point B in \Fig{fig2}(b).
At the same low temperature of $T=0.4\Delta$ but having $\epsilon_1$ shifted down to $-10\Delta$ (line A), the broad non-Kondo peak at $\omega>0$ shifts to red. This elevates the blue side of the Kondo peak substantially, resulting in a positive $S$ at point A.
These analyses affirm that the sign of $S$ is not solely determined
by the Kondo resonance. In particular, the sign reversal of $S$ from B to A in \Fig{fig2}(b) is mainly caused by the tails of non-Kondo peaks near the chemical potential as $\epsilon_1$ varies.

\begin{figure}
  \includegraphics[width=\columnwidth]{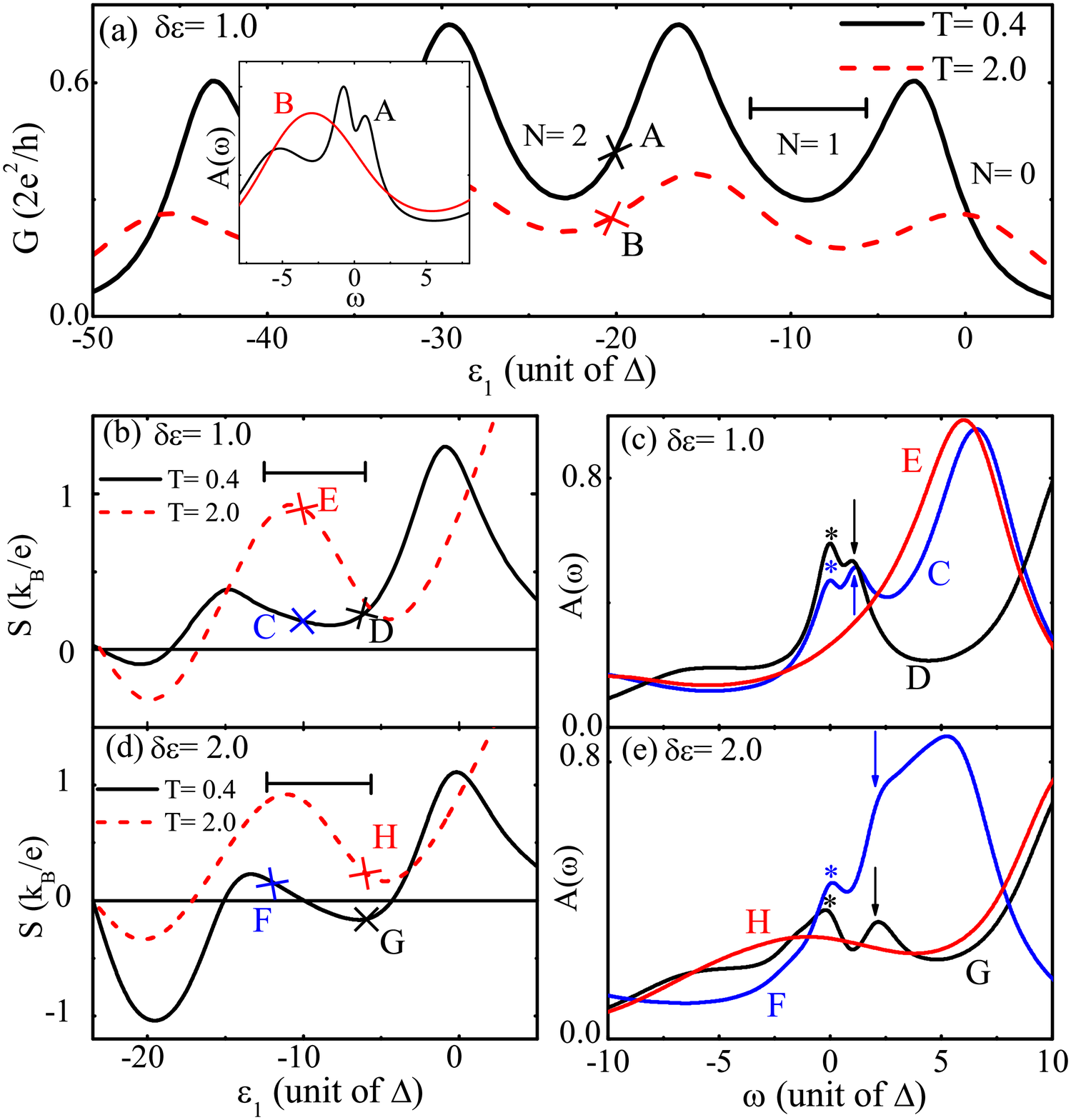}
  \caption{(Color online).
  (a) $G$ and (b) $S$ versus $\epsilon_1$ for $\delta\epsilon=\Delta$ at high and low temperatures. The lines in (c) and inset of (a) are $A(\w)$ for a number of points marked by the crosses in (a) and (b).
  Panels (d) and (e) are analogous to (b) and (c), respectively; but they are for the case of $\delta\epsilon = 2\Delta$. The horizontal bars indicate the Kondo regimes. The asterisks and vertical arrows in (c) and (e) indicate the positions of Kondo peaks and side peaks, respectively. Other parameters  (in unit of $\Delta$) are $U = 15$ and $W = 60$.
  }\label{fig3}
\end{figure}

At a low $T$, the non-Kondo spectral peaks affecting the value of $S$ have significant components from the ``side'' peaks at $\w = \mu \pm\,\delta\epsilon$ \cite{Poh97189}, whose emergence is one of the main features of multi-level effect. The side peaks originate from a two-electron resonant process, where an electron transfers from a dot level to a lead, while another electron comes in and occupies another level \cite{Ino9314725}.
In \Fig{fig2}(c) the positions of side peaks for $\epsilon_1=-10\Delta$ (line A) are indicated by vertical arrows (see Supplemental Material for details \cite{Sup1}).
The side peaks do not correspond to quasi-particle resonances \cite{Kro0369}
and survive beyond the Kondo temperature \cite{Ino9314725,Sak862231}.
Therefore, if $\delta\epsilon$ is sufficiently small, the side peaks will appear within the energy window that is critical to $S$. This is expected to lead to novel characteristics for thermopower.

We then set the inter-level spacing a relatively small value, $\delta\epsilon=\Delta$.
Figure~\ref{fig3}(a) depicts $G$ versus $\epsilon_1$ at the specified two values of $T$. It is surprising to see $G$ increases with decreasing $T$ in the $N = 2$ valley centered at $\epsilon^{\rm eh}$, where is supposed to be a non-Kondo regime.
To understand such an unconventional behavior, the corresponding $A(\w)$ are shown in the inset. At the low $T$, a pair of side peaks emerge at $\w = \mu \pm\,\delta\epsilon$, leading to the enhanced $G$ in the non-Kondo regime.

The $S$ versus $\epsilon_1$ are displayed in \Fig{fig3}(b) for $\epsilon_1 > \epsilon^{\rm eh}$. At both low and high $T$, $S$ retains positivity in the Kondo regime.
This is distinctly different from the case of $\delta\epsilon = 7\Delta$ shown in \Fig{fig2}(b), and can be understood by looking into the $A(\w)$ plotted in \Fig{fig3}(c). At the low $T$ (lines C and D), beside the Kondo peak centered at $\omega= \mu$, there exists a prominent side peak at $\w = \mu + \delta\epsilon$, which contributes predominantly to the positivity of $S$.
Moreover, as $\epsilon_1$ increases towards $\mu$ (from point C to D in \Fig{fig3}(b)), the side peak becomes more accentuated. This is consistent with the finding of Ref.\,[\onlinecite{Ino9314725}].
At the high $T$, both the Kondo and side peaks vanish, and it is the nearest Hubbard peak above $\mu$ that leads to the positive $S$ (line E).

It is thus interesting to know what would happen at an intermediate inter-level spacing. We set $\delta\epsilon=2\Delta$, and the resulting $S(\epsilon_1)$ is qualitatively different between high and low temperatures.
In \Fig{fig3}(d) $S$ shows a sign reversal in the Kondo regime
at the low $T$, while it retains positivity at the high $T$.
While the high-$T$ behavior is similar to the case of $\delta\epsilon=\Delta$, the low-$T$ scenario is more complex, as the side peak at $\omega=\mu + \delta\epsilon$ is now at the edge of the energy window; see \Fig{fig3}(e).
The sign of $S$ is thus determined by a competition among several peaks of different origins. For instance, the overlap of Kondo and Hubbard peaks at $\w < \mu$ gives rise to a negative $S$ for the line G, while the overlap of side peak and Hubbard peak at $\w > \mu$ gives a positive $S$ for the line F.

\begin{figure}
  \includegraphics[width=\columnwidth]{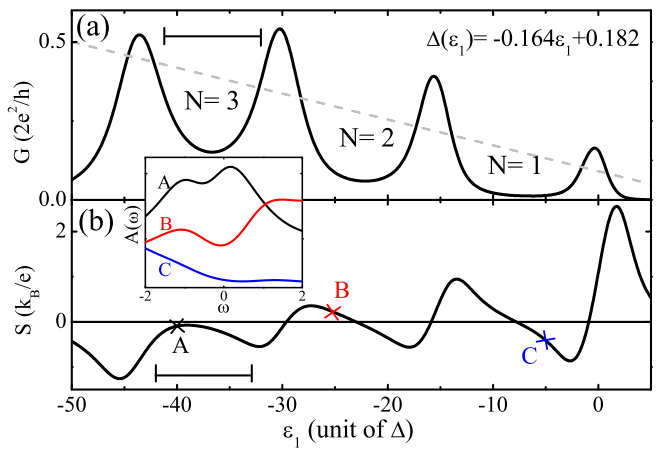}
  \caption{(Color online).
  (a) $G$ and (b) $S$ versus $\epsilon_1$ for $\delta\epsilon = \Delta$ at $T=0.4\Delta$. The linear decrease of $\Delta(\epsilon_1)=-0.164\epsilon_1+0.182$ (dashed line) is also shown in (a). The inset shows $A(\w)$ for three points marked by the crosses in (b). The horizontal bar indicate the Kondo regime. Other parameters  (in unit of $\Delta$) are $U = 15$ and $W = 60$.
  }\label{fig4}
\end{figure}

To mimic realistic experimental conditions \cite{Sch05176602}, we let the dot-lead coupling vary linearly with gate voltage (or $\epsilon_1$).
From $G$ versus $\epsilon_1$ displayed in \Fig{fig4}(a), we see the QD is tuned from the Kondo regime into the Coulomb blockade regime. In the Kondo valley of $N = 3$, $S$ retains a negative sign; see \Fig{fig4}(b). This is due to the prominent side peak at $\w = \mu -\delta\epsilon$ in $A(\w)$ (line A).
In contrast, in the non-Kondo valley of $N = 2$, both side peaks at $\w = \mu \pm\,\delta\epsilon$ are present (line B), and their competition leads to a sign reversal within the valley.
In the $N=1$ valley, $G \approx 0$ due to the Coulomb blockade \cite{Mei913048}. Both the Kondo and side peaks vanish from $A(\w)$, and the sign reversal of $S$ in this valley is caused by the competition between Hubbard peaks.

The line shape of $S$ in \Fig{fig4}(b) is overall similar to the experimental finding of Ref.\,[\onlinecite{Sch05176602}]. The major difference is that $S$ is always positive in the Kondo regime of Ref.\,[\onlinecite{Sch05176602}], while it is negative here. This may be because the $N = 3$ valley locates to the left of $\epsilon^{\rm eh}$, and it is the lower side peak at $\w = \mu -\delta\epsilon$ that dominates. It is thus expected if the QD consists of more than three levels, the $S$ would be all positive in the Kondo valley of $N=3$.

To conclude, the thermopower of a QD is determined by its spectral function near around the chemical potential. Besides Kondo correlations, the multi-level effects are also crucial to the value of thermopower.
An unconventional sign reversal behavior is predicted for QDs with small inter-level spacings at a low temperature.
Our finding and understanding shed new lights on the thermoelectric properties of strongly correlated QDs, and may be useful for the design of novel devices.

\acknowledgments

The support from the NSF of China (No.\,21103157, No.\,21033008, No.\,21233007),
the Fundamental Research Funds for Central Universities of China (No.\,2340000034, No.\,2340000025),
the Strategic Priority Research Program (B) of the CAS (XDB01020000),
and the Hong Kong UGC (AoE/P-04/08-2) and RGC (No.\,605012) is gratefully appreciated. XZ thanks Prof. M. Di~Ventra for stimulating and useful discussions.


\end{document}